\documentstyle[aps,pre,multicol,epsfig,amssymb,amsmath]{revtex}

\input{epsf}

\begin{document}

\date{\today}
\title{Toy models of crossed Andreev reflection}

\author{
R. M\'elin\thanks{melin@polycnrs-gre.fr}$^{(1)}$,
H. Jirari$^{(1)}$
and S. Peysson$^{(2)}$}
\address{$^{(1)}$Centre de Recherches sur les Tr\`es Basses
Temp\'eratures (CRTBT)\thanks{U.P.R. 5001 du CNRS,
Laboratoire conventionn\'e avec l'Universit\'e Joseph Fourier}\\
{CNRS BP 166X, 38042 Grenoble Cedex, France}}
\address{$^{(2)}$Instituut voor Theoretische Fysica,
Universiteit van Amsterdam, \\
Valckenierstraat 65, 1018XE Amsterdam, The Netherlands}
\maketitle

\begin{abstract}
We propose toy models of crossed Andreev reflection
in multiterminal hybrid structures containing out-of-equilibrium
conductors. We apply the
description to two possible experiments: (i) to a 
device containing a large quantum dot inserted in a
crossed Andreev reflection circuit. (ii) To a device
containing an Aharonov-Bohm loop inserted in a crossed
Andreev reflection circuit.
\end{abstract}

\centerline{PACS numbers: 74.80.Fp, 72.10.Bg}
\bigskip
\begin{multicols}{2}
\narrowtext


\section{Introduction}
Transport of correlated pairs of electrons
in solid state devices has focussed an important interest
recently.
Because
of the progress in the fabrication of
nanoscopic devices it will be possible
to realize in a near future transport experiments
in which two normal metal or spin polarized
electrodes are connected
to a superconductor within a distance smaller
than the BCS coherence length. 
Transport theory of multiterminal hybrid
structures has been investigated recently
with various methods (a scattering approach
in Ref.~\cite{Feinberg}, lowest order perturbation
in Ref.~\cite{Falci}, Keldysh formalism
in Ref.~\cite{Melin-Feinberg}).
Multiterminal hybrid structures
can be used to manipulate
entangled pairs of electrons in solid state
devices~\cite{Feinberg,Falci,Melin-Feinberg,Martin,Loss,Melin,tele,V_Bouchiat}
and propose new tests of quantum mechanics with
electrons such as EPR~\cite{Martin,Loss,V_Bouchiat}
or quantum teleportation experiments~\cite{tele}.
It has been suggested recently~\cite{Giroud} that proximity effect
experiments at a ferromagnet~/ superconductor interface
could be explained by spatially separated Cooper
pairs in which the spin-up (spin-down) electron
propagates in a spin-up (spin-down) domain.
This gives a strong motivation for investigating
new situations involving spatially separated
pairs of electrons, which we do in this article.
One of the possible experiments that we propose
involves a large quantum
dot inserted in a crossed Andreev reflection circuit.
This device can be used to probe spin accumulation
related to crossed Andreev reflection and elastic
cotunneling.
Another device that we propose involves an Aharonov-Bohm
loop inserted in a crossed Andreev reflection circuit.
This device can be used to probe Aharonov-Bohm oscillations
associated to spatially separated pairs of electrons.
By crossed Andreev reflection we mean the possibility that
a spin-up electron from a spin-up ferromagnetic electrode
is Andreev reflected as a spin-down hole in another spin-down
ferromagnetic electrode on the condition that the distance 
between the ferromagnetic electrodes is sufficiently small.
The other possible process is elastic cotunneling in which
a spin-up electron from a spin-up ferromagnetic electrode
is transfered as a spin-up electron in another ferromagnetic
spin-up electrode. With partially polarized ferromagnet,
crossed Andreev reflection can take place in the parallel
alignment and elastic cotunneling can take place in the
antiparallel alignment.

The theoretical description is based on
toy models relying on a series of simplifying assumptions.
The initial electrical circuit
is replaced by nodes interconnected by tunnel
matrix elements~\cite{Nazarov}.
Each node corresponds to a large quantum
dot so that the energy levels form a continuum within
each node. 
We suppose that the propagators within a given
node are uniform in space. 
A ferromagnetic
node is thus characterized the
spin-up and spin-down density
of state.
A superconducting node is characterized by
the ordinary and anomalous propagators.
We suppose that the applied voltages are small
compared to the superconducting gap and
that the propagators
at a given node are independent on energy.
Each node is supposed to be
in local equilibrium so that the distribution
function within each node is represented by 
the Fermi-Dirac distribution.
To impose Kirchoff laws we determine the spin-up
and spin-down chemical potentials so that
the spin-up and spin-down currents are conserved
at each node~\cite{Nazarov}.

\begin{figure}[tb]
\epsfxsize=5cm
\centerline{\epsfbox{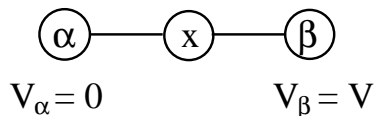}}
\medskip
\caption{The toy model used to discuss the interplay
between elastic cotunneling and sequential tunneling.
The local chemical potential at node~$x$ is determined in
such a way that current is conserved.}
\label{fig:circuit1}
\end{figure}

\section{M/M/M junction}
Let us first
consider the metal~/ metal~/ metal
junction on Fig.~\ref{fig:circuit1}
in which we suppose that a large
quantum dot (node~$x$) is inserted in between two metallic
electrodes represented by nodes~$\alpha$ and~$\beta$.
We first calculate
the Green's functions $G_{i,j}$ of the connected system
that are the solution of the chain of
Dyson equations given by
$G = g + g \otimes \Sigma \otimes G$ in a compact
notation~\cite{CCNSJ},
where $\Sigma$ is the self energy that contains
all couplings of
\begin{equation}
\label{eq:W}
{\cal W} = t_{\alpha,x} c_\alpha^+ c_x + t_{x,\alpha} c_x^+ c_\alpha
+t_{\beta,x} c_\beta^+ c_x + t_{x,\beta} c_x^+ c_\beta
.
\end{equation}
The coupling Hamiltonian (\ref{eq:W}) is formally equivalent to a 
tunnel Hamiltonian but similarly to Ref.~\cite{Cuevas} we use
a non perturbative
method based on Ref.~\cite{CCNSJ} that is valid also for
large interface transparencies. 
This method avoids inconsistencies
present in the tunneling Hamiltonian model in the regime of high 
transparency interfaces~\cite{Burstein}.
The symbol $\otimes$ in the Dyson equation includes a convolution over time
arguments and a summation over the labels of the network.
Since there is no magnetic flux one has
$t_{\alpha,x}=t_{x,\alpha}$ and $t_{x,\beta}=t_{\beta,x}$.
To obtain the current flowing from node~$\alpha$
to node~$x$
we evaluate the Keldysh Green's functions given by
$G^{+,-} = (1+G^R \otimes \Sigma) \otimes g^{+,-} \otimes
(1+\Sigma \otimes G^A)$~\cite{CCNSJ,Cuevas}. 
The current is related to the Keldysh Green's function
by the relation~\cite{CCNSJ}
\begin{equation}
\label{eq:current-normal}
I_{\alpha,x} = \frac{e^2}{h} \int \left[t_{\alpha,x}
G^{+,-}_{x,\alpha}(\omega) -
t_{x,\alpha} G^{+,-}_{\alpha,x}(\omega)
\right] d \omega
.
\end{equation}
The current flowing from node $\alpha$ to node $x$ is
given by
\begin{eqnarray}
\label{eq:cur}
I_{\alpha,x} &=& \frac{4\pi^2 t_{\alpha,x}^2 \rho_\alpha
\rho_x} { {\cal D}^2} (\mu_x-\mu_\alpha)\\
&+& \frac{4 \pi^4 t_{\alpha,x}^2 t_{\beta,x}^2 \rho_\alpha
\rho_\beta \rho_x^2} { {\cal D}^2}
(\mu_\beta-\mu_\alpha)
,
\end{eqnarray}
with ${\cal D}=1+\pi^2 t_{x,\alpha}^2 \rho_\alpha \rho_x
+\pi^2 t_{x,\beta}^2 \rho_\beta \rho_x$.
The current flowing from node $x$ to node $\beta$ is obtained
by exchanging the labels $\alpha$ and $\beta$ and changing sign.
The chemical potential $\mu_x$ in the
intermediate region is deduced from the Kirchoff law
$I_{\alpha,x}=I_{x,\beta}$:
\begin{equation}
\label{eq:mux}
\mu_x = \frac{ t_{\alpha,x}^2 \rho_\alpha \mu_\alpha
+t_{\beta,x}^2 \rho_\beta \mu_\beta}
{t_{\alpha,x}^2 \rho_\alpha + t_{\beta,x}^2 \rho_\beta}
.
\end{equation}
As expected if $t_{\alpha,x}=t_{\beta,x}$ and if
$\rho_\alpha=\rho_\beta$ we obtain
$\mu_x=\frac{1}{2}(\mu_\alpha+\mu_\beta)$.
The conductance
$G=d I / d V (V=0)$ is obtained by replacing the value
of the chemical potential (\ref{eq:mux}) in the 
expression of the current Eq.~(\ref{eq:cur}):
\begin{eqnarray}
\label{eq:current-3site}
G&=& \frac{e^2}{h} \frac{4 \pi^4 t_{\alpha,x}^2 t_{\beta,x}^2
\rho_\alpha \rho_\beta \rho_x^2}{ {\cal D}^2}\\
&\times&
\left[ 1 + \frac{1}{\pi^2 t_{\alpha,x}^2
\rho_\alpha \rho_x + \pi^2 t_{\beta,x}^2
\rho_\beta \rho_x} \right]
.
\end{eqnarray}
We recognize a contribution due to
elastic cotunneling and a contribution due
to sequential tunneling.

\begin{figure}[tb]
\epsfxsize=5cm
\centerline{\epsfbox{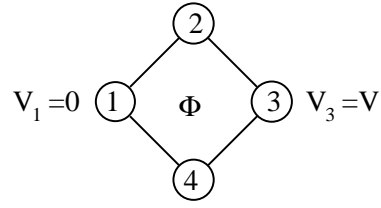}}
\medskip
\caption{The toy model used to discuss Aharonov-Bohm
oscillations.
A voltage $V_1=0$ is applied on node
$1$ and a voltage $V_3=V$ is applied on node~$3$.
The chemical potentials at nodes~$2$ and~$4$ are
determined in such a way that current is conserved.}
\label{fig:circuit2}
\end{figure}

\section{Aharonov-Bohm oscillations in normal structures}
To discuss Aharonov-Bohm oscillations
we consider the toy model on Fig.~\ref{fig:circuit2}.
We consider a symmetric structure in which
$\rho=\rho_1=\rho_3$ and
$\rho'=\rho_2=\rho_4$, and we use the notation
$u=\pi^2 t^2 \rho \rho'$. 
The conductance is given by
\begin{equation}
\label{eq:current-AB}
G = \frac{e^2}{h} \frac{4 u}{1+4 u + 2 u^2
(1-\cos{\Phi})} 
\end{equation}
which oscillates with the applied magnetic flux. 
As a consequence our
toy model can be used to describe Aharonov-Bohm
oscillations.

\section{Crossed Andreev reflection}
Now we discuss crossed
Andreev reflection~\cite{Feinberg,Falci,Melin-Feinberg}.
We consider the same toy model as on Fig.~\ref{fig:circuit1}
but now node~$x$ is superconducting.
We denote by $\rho_\alpha^\uparrow$, $\rho_\alpha^\downarrow$,
$\rho_\beta^\uparrow$ and~$\rho_\beta^\downarrow$ the
density of states of spin-up and spin-down electrons
in electrodes $\alpha$ and~$\beta$. 
We note $g$ the ordinary propagator of the superconducting
node
and $f$ the anomalous propagator. We show that $f$ and $g$
should satisfy the relation $f^2=g^2$.
Solving the Dyson equation and evaluating the Dyson-Keldysh
equation leads to the transport formula
\begin{eqnarray}
\label{eq:AR}
I^\uparrow_{\alpha,x} &=& - \frac{2 e^2}{h} V_\alpha
\frac{4 \pi^2 t_{x,\alpha}^4}{ {\cal D}^A
{\cal D}^R} f^2 \rho_\alpha^\uparrow
\rho_\alpha^\downarrow\\
\label{eq:EC}
&-&\frac{e^2}{h} \left( V_\alpha - V_\beta \right)
\frac{4 \pi^2 t_{x,\alpha}^2
t_{x,\beta}^2}{ {\cal D}^A {\cal D}^R}
\left| g^\downarrow \right|^2 \rho_\alpha^\uparrow
\rho_\beta^\uparrow \\
\label{eq:CAR}
&-&\frac{e^2}{h} \left(V_\alpha+V_\beta \right)
\frac{4\pi^2 t_{x,\alpha}^2 t_{x,\beta}^2}
{ {\cal D}^A {\cal D}^R}
f^2 \rho_\alpha^\uparrow \rho_\beta^\downarrow,
\end{eqnarray}
where the determinant ${\cal D}$ is given by
$
{\cal D}^A = 1 - i \pi g (\Gamma^\uparrow
+\Gamma^\downarrow ) + \pi^2
(f^2-g^2)\Gamma^\uparrow \Gamma^\downarrow
$, and where $\Gamma^\uparrow=|t_{x,\alpha}|^2
\rho_\alpha^\uparrow
+|t_{x,\beta}|^2\rho_\beta^\uparrow$ and
$\Gamma^\downarrow=|t_{x,\alpha}|^2\rho_\alpha^\downarrow
+|t_{x,\beta}|^2\rho_\beta^\downarrow$.
The function~$g^\downarrow$
is given by $g^\downarrow=g+i\pi (f^2-g^2)\Gamma^\downarrow$.
The term~(\ref{eq:AR})
corresponds to local Andreev
reflections, the term~(\ref{eq:EC}) corresponds
to elastic cotunneling and the term~(\ref{eq:CAR})
corresponds to crossed Andreev reflection.
In microscopic models it is possible to show that
because of averaging between the different
conduction channels the conductance associated to crossed
Andreev reflection is equal to the conductance
associated to elastic cotunneling~\cite{Falci,Melin-Feinberg}.
This cannot be demonstrated in our toy model
because we lost all information about the spatial
dependence of the propagators in the superconductor.
Instead we choose the
propagators $f$ and $g$ in such a way that
the conductance associated to elastic
cotunneling is identical to the conductance
associated to crossed Andreev reflection.
This leads to the condition $f^2=g^2$.
This condition ensures
the equality
of the Andreev reflection and elastic cotunneling
currents also for large interface transparencies.

\section{Effect of an out-of-equilibrium conductor}
\begin{figure}[tb]
\epsfxsize=6cm
\centerline{\epsfbox{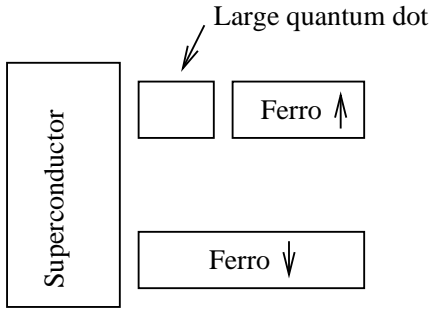}}
\medskip
\caption{Geometry of the device in which a large quantum dot
is inserted in a crossed Andreev reflection circuit.}
\label{fig:dev1}
\end{figure}

\begin{figure}[tb]
\epsfxsize=6cm
\centerline{\epsfbox{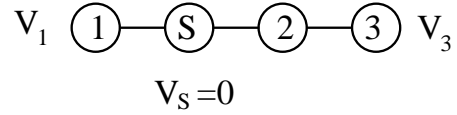}}
\medskip
\caption{The toy model corresponding to Fig.~\ref{fig:dev1}.
Node~$1$ corresponds to the spin-down electrode 
on Fig.~\ref{fig:dev1}. Node~$3$ corresponds to the
spin-up electrode and node~$2$ corresponds to the large
quantum dot.
}
\label{fig:circuit6}
\end{figure}

\begin{figure}[tb]
\epsfxsize=8cm
\centerline{\epsfbox{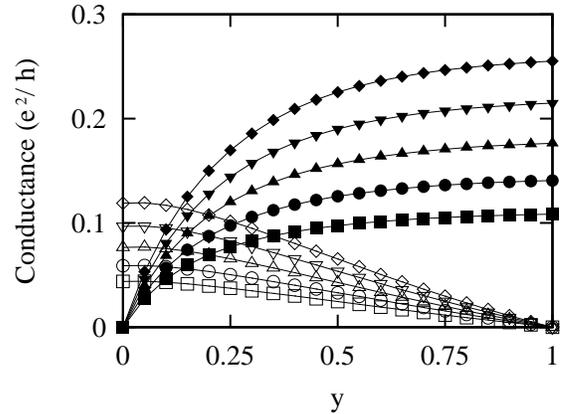}}
\medskip
\caption{Comparison between the crossed conductances
in two situations for the toy model on Fig.~\ref{fig:circuit6}:
(i) $V_1=0$ and $V_3=V$
(open symbols);
and (ii) $V_1=V$ and $V_3=0$ (filled symbols).
The density of states are given by
$\rho_N=1$,
$\rho_{1,\uparrow}=1$, $\rho_{1,\downarrow}=0$,
$\rho_{2,\uparrow}=y$, $\rho_{2,\downarrow}=1$,
$\rho_{3,\uparrow}=y$ and $\rho_{3,\downarrow}=1$.
The different curves correspond to different values of
the hopping parameter:
$t=0.11$ ($\Box$ and $\blacksquare$);
$t=0.12$ ($\circ$ and $\bullet$);
$t=0.13$ ($\Delta$ and $\blacktriangle$);
$t=0.14$ ($\triangledown$ and $\blacktriangledown$);
$t=0.15$ ($\lozenge$ and $\blacklozenge$).
}
\label{fig:G-accu}
\end{figure}

Now we discuss a device containing a large quantum dot
inserted in a crossed Andreev reflection circuit.
The non equilibrium
effects are described by the out-of-equilibrium distribution
function $n_F(\omega-\mu_{i,\sigma})$ where $\mu_{i,\sigma}$
is equal to the local chemical potential of spin-$\sigma$
electrons in node $i$. The values of $\mu_{i,\sigma}$ are
determined in such a way that current is conserved.
The geometry of the device is shown on
Fig.~\ref{fig:dev1} and
the toy model is shown on Fig.~\ref{fig:circuit6}.
Without the quantum dot the current response is
symmetric: for the toy model on Fig.~\ref{fig:circuit1}
with node~$x$ being superconducting
the crossed conductance ${\cal G}_{\alpha,\beta}
=\partial
I_\alpha (V_\alpha,V_\beta) 
/ \partial V_\beta$ is equal to
${\cal G}_{\beta,\alpha}=\partial
I_\beta (V_\beta,V_\alpha)
/ \partial V_\alpha$~\cite{Falci,Melin-Feinberg}.
As it is visible on Fig.~\ref{fig:G-accu}
such symmetry relations are not valid
for the device on Fig.~\ref{fig:dev1}.
With the parameters on Fig.~\ref{fig:G-accu} we see that
the total density of states at nodes $2$ and $3$ is
equal to $1+y$. The absolute value of the
density of states $\rho$ can be larger than unity.
Generically
if the densities of states increase at one point we will be
out of the convergence radius of the perturbative series
(namely the term of order $t^4$ becomes larger than the term of
order $t^2$,
the term of order $t^6$ becomes larger than
the term of order $t^4$, etc)
but the values of the tunnel matrix elements
are small enough so that we are not in
this regime.
The limiting cases $y=0$ and $y=1$ can be understood on simple
grounds (see Fig.~\ref{fig:G-accu}).
For $y=0$ node~$1$ is a half-metal spin-up ferromagnet
(meaning that only spin-up electrons are present in node~$1$)
and nodes~$2$ and~$3$ are half-metal spin-down ferromagnets.
If we suppose that $V_1=V \ne 0$ and $V_3=V_S=0$ we
see that there is no voltage difference between
the superconductor and node~$3$. As a consequence the
spin-down electrons arising from crossed Andreev
reflection cannot be transmitted to node~$3$ and
give rise to charge accumulation at node~$2$.
There is no current in this case.
If we suppose that $V_1=0$, $V_3=V \ne 0$ and $V_S=0$ we
see that spin-down electrons arising from crossed Andreev
reflection can be transmitted to node~$3$ because
there is a voltage difference between the superconductor
and node~$3$. There is a finite current, in agreement
with Fig.~\ref{fig:G-accu}.
Another limiting case is $y=1$ (see Fig.~\ref{fig:G-accu}).
For $y=1$ node~$1$ is
a half-metal spin-up ferromagnet, nodes~$2$ and~$3$ are
normal metal. We find that the crossed
conductance is finite if $V_1=V\ne 0$, $V_3=0$ and 
is zero if $V_1= 0$ and $V_3=V\ne 0$ (see Fig.~\ref{fig:G-accu}).
This should be contrasted with the case $y=0$ where the opposite
occurs.
The case 
$V_1=0$ and $V_3=V\ne 0$ can be understood from
a cancellation between the currents due to
crossed Andreev reflection and elastic cotunneling
so that the crossed conductance is zero unlike the
case $y=0$ discussed above.
The case $V_1=V\ne 0$, $V_3=0$ is more complex
because it involves spin accumulation at node~$2$
in the presence of elastic cotunneling and crossed
Andreev reflection. Spin-up electrons from node~$2$
are transferred to node~$1$ because of elastic
cotunneling. Because of crossed
Andreev reflection spin-up electrons are transferred
to node~$1$ and spin-down electrons are transferred
to node~$2$. The rate of the two processes is
identical and there is thus no charge accumulation
like in the case $y=0$ discussed above.
A naive argument would suggest
that the spin-up and spin-down chemical potentials
at node~$2$ take opposite values and that no charge
current is flowing from node~$2$ to node~$3$ but only
a spin current is flowing. As
seen on Fig.~\ref{fig:G-accu} this is not the case
since we find a finite charge current 
flowing from node~$2$ to node~$3$. 
We suggest that
the naive reasoning does not work because the distribution
function at node~$2$ may be affected by the fact that the
two electrons arising from crossed Andreev reflection
have an opposite energy whereas elastic cotunneling
is at constant energy. This may induce a non trivial
distribution function at node~$2$.

\begin{figure}[tb]
\epsfxsize=6cm
\centerline{\epsfbox{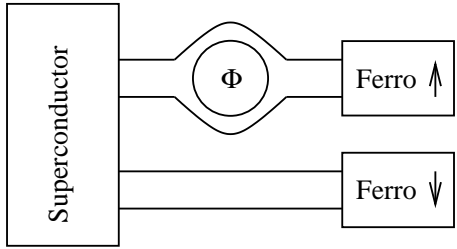}}
\medskip
\caption{Geometry of the Aharonov-Bohm experiment.}
\label{fig:AB}
\end{figure}
\begin{figure}[tb]
\epsfxsize=6cm
\centerline{\epsfbox{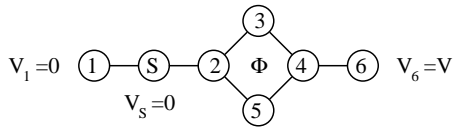}}
\medskip
\caption{The toy model used to discuss the Aharonov-Bohm
experiment represented on Fig.~\ref{fig:AB}.}
\label{fig:AB-circuit}
\end{figure}

\section{Aharonov-Bohm effect related to pair states}
\begin{figure}[tb]
\epsfxsize=8cm
\centerline{\epsfbox{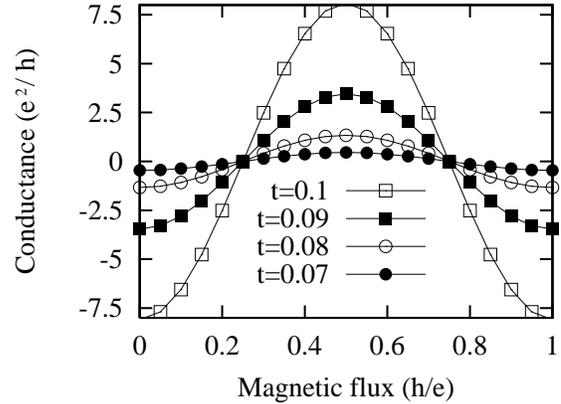}}
\medskip
\caption{Aharonov-Bohm effect related to non
separable correlations. Node~$1$ is a half-metal
spin-up ferromagnet and the Aharonov-Bohm loop
is a half-metal spin-down ferromagnet:
$\rho_{i,\uparrow}=1$ and
$\rho_{i,\downarrow}=0$,
$i=2, ... , 6$.
We have shown the
variation of $G_{S,1}(\Phi)-G^{\rm av}_{S,1}$
as a function
of $\Phi$. The average conductance defined as
$G^{\rm av}_{S,1}=(G_{S,1}(\Phi=0)+
G_{S,1}(\Phi=\pi))/2$ is
positive. 
The tunnel amplitudes are identical for
all links of the network:
$t=0.1$ ($\Box$);
$t=0.09$ ($\blacksquare$);
$t=0.08$ ($\circ$);
$t=0.07$ ($\bullet$). We obtain 
the same oscillations with a parallel
spin orientation of the ferromagnetic
electrodes but with an opposite current.
} 
\label{fig:AB-half}
\end{figure}

Now we consider the device on Fig.~\ref{fig:AB} in
which an Aharonov-Bohm loop is inserted in a crossed
Andreev reflection circuit and
we use the toy model on Fig.~\ref{fig:AB-circuit}.
We first consider that the Aharonov-Bohm loop is
a half-metal ferromagnet. In this case the only
phenomenon coming into account is crossed
Andreev reflection for antiparallel spin
orientations and elastic cotunneling for parallel
spin orientations. We obtain
Aharonov-Bohm oscillations
with a negative magnetoresistance (see
Fig.~\ref{fig:AB-half}).
It may be difficult to probe this situations in
experiments because the phase coherence length in
ferromagnetic metals is very small. This is why we
consider the case where the Aharonov-Bohm loop
is made of a normal metal.

\begin{figure}[tb]
\epsfxsize=8cm
\centerline{\epsfbox{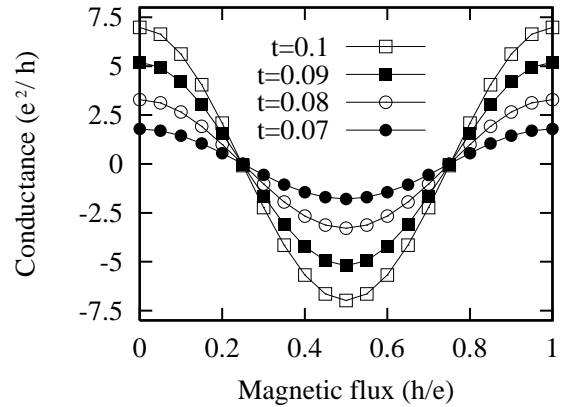}}
\medskip
\caption{The same as Fig.~\ref{fig:AB-half} but now
nodes $2,..,5$ are unpolarized:
$\rho_{i,\uparrow}=\rho_{i,\downarrow}=1$
with $i=2,...,5$.
} 
\label{fig:AB-EPR}
\end{figure}

Several phenomena come into account:
(i) crossed Andreev reflection; (ii)
elastic cotunneling; (iii) spin accumulation 
in the Aharonov-Bohm loop (the local spin-up and
spin-down chemical potentials are not equal);
(iv) electroflux effect similar to Ref.~\cite{Stoof}
(the local chemical
potential oscillates with the magnetic
flux applied on the Aharonov-Bohm loop).
Another possible process occurring in this structure 
with highly transparent 
interfaces is that the two electrons
of a Cooper pair are transfered into the Aharonov-Bohm loop,
couple to the magnetic flux and come back in the other
electrode.
As it is visible on Fig.~\ref{fig:AB-EPR} we obtain
also Aharonov-Bohm oscillations
but with a positive magnetoresistance.

\section{Conclusion}

To summarize we have provided a toy model
of multiterminal hybrid structures containing
large quantum dots. We applied the
model to a device in which a large quantum dot is
inserted in a crossed Andreev reflection circuit.
We also proposed another possible experiment 
intended to probe an Aharonov-Bohm effect related
to spatially separated pairs of electrons.
These two situations may be the object of
future experiments. The description was
based on a toy model in which the
initial electrical circuit is replaced by a set of nodes
interconnected by tunnel matrix elements. It is expected
that the qualitative behavior is captured by our toy model.
The formulation of a microscopic theory  is left as an important
open question. Within a microscopic theory it would be possible
to discuss rigorously the role played by the elastic mean
free path and the phase coherence length. In our toy model
each node is described by a single propagator. As a consequence
we have lost all information about the spatial variation
of the propagators and we suppose implicitly that the 
distance between the contacts is smaller than the elastic
mean free path. But qualitatively for the device
on Fig.~\ref{fig:AB} the current is proportional to the
square of the anomalous propagator in the superconductor,
and to the square of the ordinary propagator in the normal
metal electrodes. We have supposed in our toy model that
the distance $D$ between the contacts on the superconductor
is much smaller than the BCS coherence length $\xi_{\rm BCS}$
and the length $R$ of the normal metal electrodes is much smaller
than the phase coherence length $l_{\phi}$. 
Qualitatively
the conductances on Figs.~\ref{fig:G-accu},~\ref{fig:AB-half}
and Fig.~\ref{fig:AB-EPR} should thus be multiplied by the
exponential factors $\exp{(-D/\xi_{\rm BCS})}$ and
$\exp{(-R/l_\phi)}$.
Another important ingredient is the role
played by interface transparencies. We could solve the
toy model with arbitrary interface transparencies and
obtained in some cases a crossed conductance larger than $e^2/h$
with large interface transparencies. As discussed previously
the value of the conductance would be much smaller in a
microscopic model because of the exponential dependence
of the propagators in the superconductor.
To gain
in realism it would be also interesting to discuss
diffusive models in connection with the occurrence
of $h/e$ oscillations in the conductance
predicted from the toy model. 
$h/e$ oscillations have been also obtained 
for the proximity effect in other
geometries (see Ref.~\cite{Fuku}).

The authors acknowledge fruitful discussions with
D. Feinberg.

\end{multicols}
\end{document}